\documentclass[useAMS, usenatbib, referee]{biom}

\def\bSig\mathbf{\Sigma}

\usepackage[figuresright]{rotating}
\usepackage{commath}
\usepackage{soul, color}
\usepackage{amsfonts}
\usepackage{multirow}
\usepackage{booktabs}
\usepackage{multicol}
\usepackage{rotating}
\usepackage{comment}
\usepackage{float}
\usepackage{graphicx}
\graphicspath{{../Images/}}
\usepackage{hyperref}
\hypersetup{colorlinks = true,
citecolor = blue,
linkcolor = blue }

\soulregister\cite7 %for highlighting within macro
\newcommand*{\Cov}{\text{Cov}}

\newcommand*{\Var}{\text{Var}}
\title[Nonparametric Estimation of the Potential Impact Fraction and the Population Attributable Fraction]{Nonparametric Estimation of the Potential Impact Fraction and the Population Attributable Fraction  with Individual-Level and Aggregated Data}

\author{
Colleen E. Chan$^{1}$,
Rodrigo Zepeda-Tello$^{2}$,
Dalia Camacho-Garc\'{i}a-Forment\'{i}$^{2}$, \\
Frederick Cudhea$^{3}$, Rafael Meza$^{4}$, Eliane Rodrigues$^{5}$, \\ Donna Spiegelman$^{6,*}\email{donna.spiegelman@yale.edu}$, Tonatiuh Barrientos-Gutierrez$^{2,**}\email{tbarrientos@insp.mx}$, and Xin Zhou$^{6,***}\email{xin.zhou@yale.edu}$\\
$^{1}$Department of Statistics and Data Science, Yale University, New Haven, Connecticut, U.S.A. \\
$^{2}$National Institute of Public Health of Mexico, Mexico \\
$^{3}$Friedman School of Nutrition Science and Policy, Tufts University, Boston, Massachusetts, U.S.A. \\
$^{4}$Department of Epidemiology, University of Michigan School of Public Health, Ann Arbor, Michigan, U.S.A. \\
$^{5}$Instituto de Matem\'{a}ticas, Universidad Nacional Aut\'{o}noma de M\'{e}xico, Mexico \\
$^{6}$Department of Biostatistics, Yale School of Public Health, New Haven, Connecticut, U.S.A.
}

\begin{document}

%\date{{\it Received October} 2004. {\it Revised February} 2005.\newline
%{\it Accepted March} 2005.}

\pagerange{\pageref{firstpage}--\pageref{lastpage}} \pubyear{2022}

\volume{59}
\artmonth{XX}
\doi{10.1111/j.1541-0420.2005.00454.x}

%  This label and the label ``lastpage'' are used by the \pagerange
%  command above to give the page range for the article

\label{firstpage}

%  pub the summary here

\begin{abstract}
The estimation of the potential impact fraction, including the population attributable fraction, with continuous exposure data frequently relies on strong distributional assumptions. However, these assumptions are often violated if the underlying exposure distribution is unknown. In this article, we discuss the impact of distributional assumptions in the estimation of the population impact fraction, showing that distributional violations lead to biased estimates. We propose nonparametric methods to estimate the potential impact fraction for aggregated data, where only the exposure mean and standard deviation are available, or individual data, where the full exposure distribution can be estimated from a sample of the target population. The finite sample performance of the proposed methods is demonstrated through simulation studies. We illustrate our methodology with a study of the impact of eliminating sugar-sweetened beverage consumption on the incidence of type 2 diabetes in Mexico. We also developed the \textsf{R} package \texttt{pifpaf} to implement these methods.
\end{abstract}

%
%  Please place your key words in alphabetical order, separated
%  by semicolons, with the first letter of the first word capitalized,
%  and a period at the end of the list.
%


\maketitle

\section{Introduction}
\label{s:intro}

The potential impact fraction (PIF), also known as the generalized impact fraction, quantifies the contribution of an exposure to disease cases, morbidity, or mortality, by estimating the  difference in the proportion of cases resulting from a change in the exposure distribution to a counterfactual scenario \citep{levin1953paf, murray2003comparative, walter1976paf, vander2004estimating}.
It evaluates the burden of disease that would be prevented in a population if the exposure were to be shifted to some counterfactual exposure level. 
A special case of the PIF is the population attributable fraction (PAF), often referred to as the population attributable risk, where the counterfactual exposure equals the baseline level, which in many cases is no exposure, for all individuals. For instance, in our illustrative example, we estimate the proportion of type 2 diabetes cases that would be prevented if no one consumed any sugar-sweetened beverages. 
While the selection of counterfactuals has focused on public health scenarios, little has been discussed about the methodological implications of parametric distributional assumptions on the accuracy and reliability of PIF estimation \citep{murray2003comparative}.

In what we call the ``standard method'', the PIF is estimated as a function of the mean and variance of the exposure obtained from national surveys \citep{gortmaker2015cost, danaei2010promise, maredza2016burden, lawes2006blood, gmel2011estimating, Veerman2016australians} and meta-analytical relative risks \citep{murray2003comparative, forouzanfar2015global}. It is assumed that the exposure follows a specific probabilistic distribution, usually without empirical verification, and that the same probabilistic distribution applies to different settings and countries, as implemented, for example, in the case of the Global Burden of Disease project \citep{forouzanfar2015global}. % page 6
 The implications of potential mis-specification of these distributions are rarely discussed, despite previous evidence showing that the PIF estimates can be biased if the exposure distribution is misspecified \citep{kehoe2012alcohol}.

The problem of model misspecification when estimating the PIF is not new. Both semiparametric and nonparametric methods have been proposed to avoid making distributional assumptions in cases where both the relative risk and the exposure distribution are obtained from the same cohort \citep{chen2010attributable, sjolander2014doubly, taguri2012doubly}. However, international efforts to estimate the burden of disease in different countries, such as those conducted by the World Health Organization and the Institute for Health Metrics and Evaluation, are usually based on relative risks from meta-analyses and population-level distributional parameter estimates from survey data, in which case nonparametric methods are not available \citep{forouzanfar2015global}. Even in cases where disaggregated individual-level exposure data are available, the lack of methods to estimate the PIF by combining these data with meta-analytical relative risks has forced researchers to aggregate the data and follow the standard method \citep{Veerman2016australians, gortmaker2015cost}.

In this paper, we investigate the problems that arise when the PIF is estimated through an arbitrary selection of the exposure distribution. We then propose two nonparametric methods that estimate the PIF using meta-analytical relative risks that largely avoids the problems identified, the first of which uses individual-level exposure data and the second of which only uses the mean and variance of the exposure data. Finally, we conduct numerical simulations to evaluate the performance of our proposed methods and illustrate them in a study of the impact of reducing sugar-sweetened beverage consumption on the incidence of type 2 diabetes in Mexico.

% --------------------------------------------------------------------------------------------------------------------------

\section{Background}

\subsection{Potential Impact Fraction}
\label{seq:PIFdef}

Let $\bm{X} = (X_1, X_2, \dots, X_k)^T$ with $k$ elements be the exposure of interest, which takes values over $\mathcal{X}$, the set of all possible $\bm{X}$. 
Here, we take $\bm{X}$ as a vector instead of a scalar to incorporate general situations where there may exist multiple exposure variables of interest or where the exposure may be a categorical variable with multiple levels. The estimation of the PIF requires a relative risk function, $RR$, that depends on the exposure $\bm{X}$, and $\bm{\beta} = (\beta_0, \beta_1, \dots, \beta_{k})^T$, the regression coefficients corresponding to the exposure $\bm{X}$ that are usually obtained from a previous study or a meta-analysis. Here, we assume that $\bm{\beta}$ is a causal parameter if all confounders were adjusted for in the regression model. Examples of relative risk functions include the \textit{exponential} function, $RR(\bm{X};\bm{\beta}) = \textrm{exp}(\bm{\beta}^T \bm{X})$, for logistic, Poisson, or Cox regression models, and the \textit{linear} function, $RR(\bm{X};\bm{\beta}) = 1 + \bm{\beta}^T \bm{X}$, for linear regression models.

When the exposure is categorical, the PIF is defined as
\begin{equation}\label{discretepif}
\textrm{PIF} =
\dfrac{\sum_{\bm{X} \in \mathcal{X}} p_{\textrm{obs}}(\bm{X}) \cdot RR(\bm{X};\bm{\beta}) - \sum_{\bm{X} \in \mathcal{X}}  p_{\textrm{cft}}(\bm{X}) \cdot RR(\bm{X};\bm{\beta})}{\sum_{\bm{X} \in \mathcal{X}}  p_{\textrm{obs}}(\bm{X}) \cdot RR(\bm{X};\bm{\beta})},
\end{equation}
where $p_{\textrm{obs}}(\bm{X})$ is the observed probability mass function of the exposure $\bm{X}$ in the population to which the PIF will be applied, and $p_{\textrm{cft}}(\bm{X})$ is the probability mass function in the counterfactual scenario.
Alternatively, if $\bm{X}$ is continuous,  the PIF is given by
\begin{equation}\label{continuouspif}
\textrm{PIF} = \dfrac{\int_{\mathcal{X}} RR(\bm{X};\bm{\beta})f_{\textrm{obs}}(\bm{X})\text{d}\bm{X} - \int_{\mathcal{X}} RR\big(\bm{X};\bm{\beta} \big)f_{\textrm{cft}}(\bm{X}) \text{d}\bm{X}}{\int_{\mathcal{X}} RR(\bm{X};\bm{\beta})f_{\textrm{obs}}(\bm{X})\text{d}\bm{X}},
\end{equation}
where $f_{\textrm{obs}}$ and $f_{\textrm{cft}}$ represent probability density functions of the observed exposure and the counterfactual continuous exposure, respectively \citep{murray2003comparative, vander2004estimating}.

In general, unifying \eqref{discretepif} and \eqref{continuouspif} to allow for both discrete and continuous exposures, the PIF can be written as:
\begin{equation}\label{pifdef}
\textrm{PIF} = \dfrac{ \mathbb{E}^{\textrm{obs}}_{\bm{X}} \Big[  RR\big(\bm{X};\bm{\beta} \big)\Big] - \mathbb{E}^{\textrm{cft}}_{\bm{X}} \Big[  RR\big(\bm{X};\bm{\beta}\big)\Big]}{\mathbb{E}^{\textrm{obs}}_{\bm{X}} \Big[  RR\big(\bm{X};\bm{\beta} \big)\Big]},
\end{equation}
where  $\mathbb{E}^{\textrm{obs}}_{\bm{X}} \left[  RR(\bm{X};\bm{\beta})\right]$ represents the expected value of the relative risk under the observed exposure distribution in a given population  and $\mathbb{E}^{\textrm{cft}}_{\bm{X}} \left[  RR(\bm{X};\bm{\beta})\right]$ is the expected value of the relative risk under a counterfactual distribution of the exposure \citep{taguri2012doubly, wang2012comparative}.

Often, the counterfactual exposure distribution can be represented as a transformation $g$ on the exposure $\bm{X}$. For example, $g(\bm{X}) = 0.6 \cdot \bm{X}$ might represent a $40 \%$ reduction in the exposure, or $g(\bm{X}) = \bm{X} - 2$, an overall decrease of $2$ units of the exposure. Then, the PIF can be written as
\begin{equation}\label{eqpifeq}
\textrm{PIF} = \dfrac{ \mathbb{E}^{\textrm{obs}}_{\bm{X}} \Big[  RR\big(\bm{X};\bm{\beta} \big)\Big] - \mathbb{E}^{\textrm{obs}}_{\bm{X}} \Big[  RR\big(g(\bm{X});\bm{\beta}\big)\Big]}{\mathbb{E}^{\textrm{obs}}_{\bm{X}} \Big[  RR\big(\bm{X};\bm{\beta} \big)\Big]}.
\end{equation}

The PAF is defined as a special case of the $\textrm{PIF}$ when the counterfactual is the baseline exposure for all individuals (\textit{i.e.}, $g(\bm{X}) = \bm{0}$ and $RR\big(g(\bm{X});\bm{\beta}\big)=1)$. %$E^{\textrm{cft}}_{\bm{X}} \left[  RR(\bm{X};\bm{\beta})\right] = 1$
Thus, the expected value of the relative risk under the counterfactual scenario equals 1 \citep{vander2004estimating}, yielding
\begin{equation}\label{eq:PAF}
\textrm{PAF}
=  1 - \dfrac{1}{\mathbb{E}^{\textrm{obs}}_{\bm{X}} \Big[  RR\big(\bm{X};\bm{\beta} \big)\Big]}.
\end{equation}

%The PIF and PAF have \edit{causal interpretations} if the following assumptions hold. First, $\bm{\beta}$ is a causal parameter, which holds if the model to estimate the relative risk \edit{has no bias or confounding present}. Second, $\bm{\beta}$ is also transportable to the counterfactual population if the confounder distribution is similar in the observed population and the counterfactual confounder distribution. Third, we also assume that there are no effect modifiers between the association between the exposure and the outcome. See \cite{mansournia2018population} for more  details.

The PIF and PAF are identified as causal parameters when $\hat{\bm{\beta}}$ is a causal parameter and following the standard causal inference assumptions of exchangeability, positivity, and consistency in a survival data analysis setting, as in Equations 26-28 in Section 6 of \cite{young2020causal}, respectively. We will assume these to be true for the remainder of the paper. These assumptions are relevant to settings with competing events, including censoring, which are common in observational epidemiology, and as occurs in our Illustrative Example.

\subsection{Problems with the standard method}

When the exposure is categorical, the PIF can be easily estimated from equation \eqref{discretepif} \citep{Spiegelman2007:PAR}. PIF estimation is more challenging for continuous exposures.
The standard method assumes a distribution for the exposure $\bm{X}$ and estimates its parameters by matching the estimated mean and variance of the empirical exposure data to the assumed distribution of $\bm{X}$.  The PIF is then estimated using equation \eqref{continuouspif} through analytic, numerical integration or Monte Carlo integration. 

 This method depends heavily on the choice of the exposure distribution. The exposure is often assumed to follow a normal distribution, a log-normal distribution, a Weibull distribution, or other distributions. For example, studies estimating the $\textrm{PAF}$ of obesity-related diseases assumed that the exposure variable was log-normally distributed \citep{barendregt2010categorical, Veerman2016australians}. \cite{forouzanfar2015global} found that the normal distribution was best in fitting iron deficiency and low bone mineral density data while the log normal distribution was best for systolic blood
pressure, body-mass index (BMI), fasting plasma glucose, and cholesterol data.
%However, the \textit{true} distribution is unknown in practice.
%An incorrect distribution that misrepresents the data could yield a result that diverges substantially from the actual fraction, introducing a subjective bias.
If the true distribution of the population to which the results are to be applied, with it be the study population or some external one, diverges substantially from what was assumed when calculating the PIF or PAF, these quantities will be biased.
Consequently, approaches to estimate the PIF should avoid untested distributional assumptions about $\bm{X}$.
Table \ref{tab:relbias} shows how the bias of the PAF of the standard method changes as a function of the true exposure distribution (Gamma($k=1.15, \theta = 1.29$), Normal($\mu = 1.48, \sigma^2 = 1.38^2$), or Weibull($k = 1.08, \lambda = 1.53$)) when other distributions are assumed (Gamma, Lognormal, Normal, Weibull) and the relative risk function is exponential, $RR(\bm{X};\bm{\beta}) = \textrm{exp}(\bm{\beta}^T  \bm{X})$, where $\bm{\beta} = \log (1.27)$, which is taken from our illustrative example. The parameters of the true exposure distributions are also taken from our illustrative example.

\begin{table}[!htbp] \centering 
  \caption{Relative bias percentage of the PAF under different distributional assumptions for the standard method.} 
  \label{tab:relbias}
\begin{tabular}{@{\extracolsep{2pt}} lccccc} 
\\[-1.8ex] \toprule
 &   \multicolumn{4}{r}{ \multirow{2}{*}{{\it Distribution assumed}}} \\ \\[-1.8ex] 
\cline{3-6} \\
True distribution & True PAF & Gamma & Log normal & Normal & Weibull \\ 
\hline \\[-1.8ex] 
Gamma($k = 1.15, \theta = 1.29$) & 0.3455 & 0 & 189.4 & -19.6 & -0.2 \\ 
Normal($\mu = 1.48, \sigma^2 =1.38^2$) & 0.3795 & -9.2 & 163.5 & 0 & -9.3 \\ 
Weibull($k = 1.08, \lambda = 1.53$) & 0.3447 & 0.2 & 190.1 & -19.2 & 0 \\ 
\bottomrule \\[-1.8ex] 
\end{tabular} 
\end{table}

 When the assumed distribution is log-normal and the relative risk function is exponential, the $\textrm{PAF}$ equals $1$ for positive $\bm{\beta}$, as shown in Table \ref{tab:relbias}, since the denominator in \eqref{eq:PAF} equals the moment generating function of the log-normal distribution, which is infinite \citep{casella2002statistical}.
The problem results from the combination of a heavy-tailed distribution with an exponential relative risk. A random variable $X$ is said to have a heavy tail if the tail probabilities $P(X>x)$ decay more slowly than tails of any exponential distribution, that is,
$\displaystyle \lim_{x\rightarrow\infty} e^{cx}P(X>x)=\infty$ for all positive $c$. In addition to the log-normal distribution, the Pareto, Cauchy, and Weibull (with shape parameter less than 1) distributions are also heavy-tailed \citep{Foss2013:HeavyTail}. Heavy-tailed distributions such as the Pareto, Weibull, and log-normal distributions are often used to describe censored survival times. The PIF could be undefined with an exponential relative risk for a heavy-tailed distribution. In practice, the observed exposure is bounded so the tail is not heavy. The standard parametric method for PIF estimation falsely enlarges the contribution of the tail. Hence, additional constraints are required to correctly estimate the PIF.  This problem has been pointed out previously without much mathematical detail by \cite{kehoe2012alcohol}.

\begin{figure}[!htb]
    \centering
    \includegraphics[]{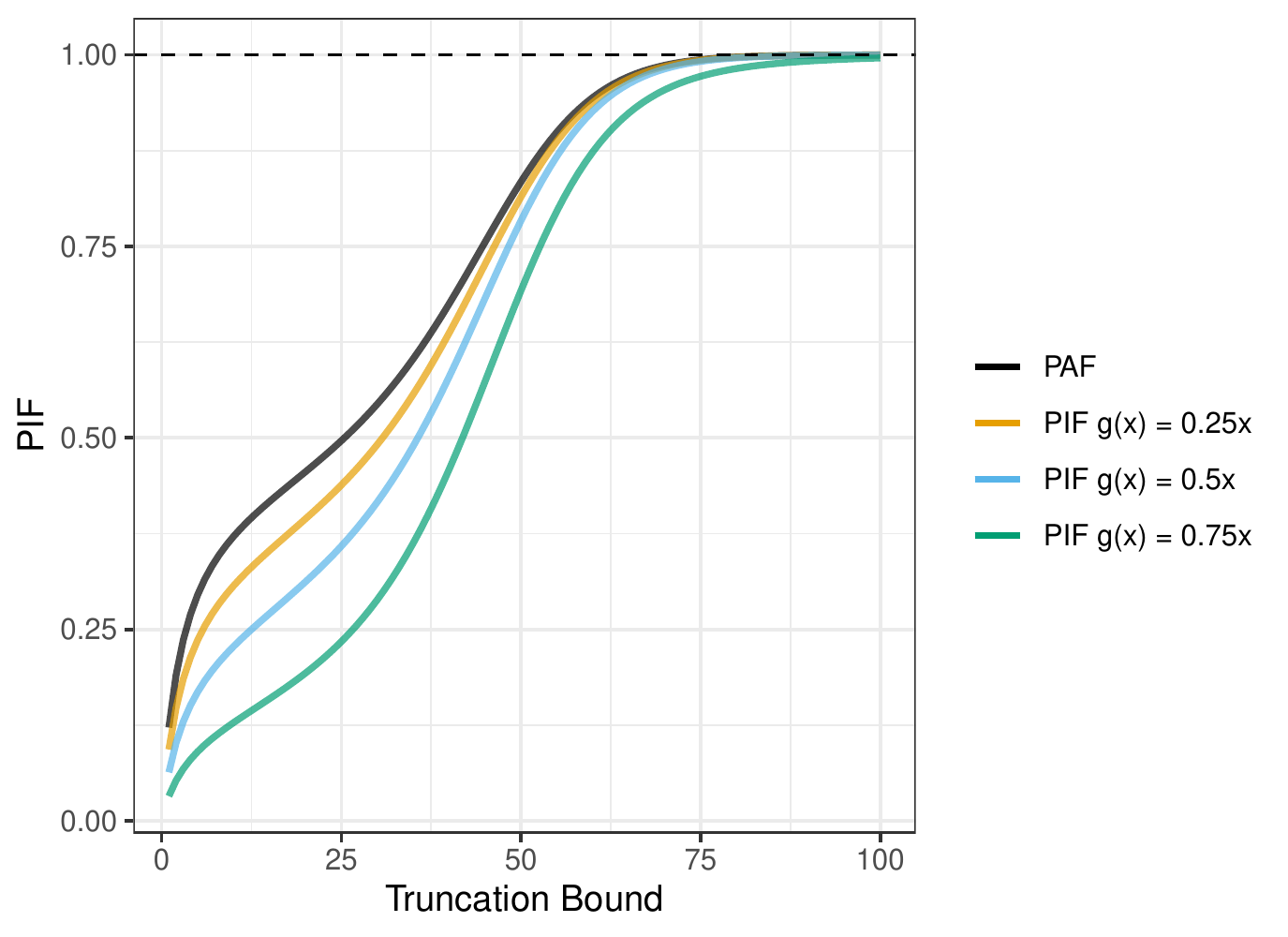}
    \caption{The PAF and three different PIFs as a function of the truncation limit $M$ assuming ${X}$ to be log-normally distributed with parameters $\log \mu=0.05,\log \sigma=0.98$ and an exponential relative risk function $RR({X};{\beta}) = e^{\beta X}$, where $\beta = \log(1.27)$. Parameters are taken from the illustrative example in Section \ref{sec:illex}. }
      \label{fig:truncated}
\end{figure}

As a potential solution to this problem, \cite{kehoe2012alcohol} truncated the assumed exposure distribution by providing an upper bound $M$, thereby avoiding large exposure values and the infinite expected relative risks. In addition, the zero (baseline exposure) and non-zero values of the exposure data are first separated, and the parameters of the positive values of the exposure are estimated using maximum likelihood estimation; thus, we refer to this method as the ``mixture method''. This can be written as
\begin{equation}
\label{eq:mixturePAF}
\textrm{PAF} 
= 1 -  \frac{1}{p_0  + (1- p_0) \int_0^M  RR(x;\beta) f(x)\text{d}x
 \big/ \int_0^M  f(x)\text{d}x} ,
\end{equation}
where $M$ is the truncation bound, and $p_0$ is the proportion of zero values in the exposure. 
Nevertheless, additional problems arise because the estimated PIF and PAF values now depend on the arbitrarily specified upper bound. For example, consider Figure \ref{fig:truncated} which shows the PAF (black) and three different PIFs as a function of the exposure's upper bound $M$. The figure shows that if an upper bound of $M = 25$ is selected, the resulting PAF is approximately $50\%$; truncating at $M = 40$ results in a PAF of $80\%$. By changing the truncation bound, $M$, we can obtain PAF estimates ranging from anywhere between $0\%$ to $100\%$. Here, we assume ${X}$ to be log-normally distributed  with parameters $\log \mu = 0.05,\log \sigma = 0.98$ and an exponential relative risk function $RR(\bm{X};\bm{\beta})$ with $\beta = \log (1.27)$. The relative risk function and fitted parameters are taken from the exposure of our illustrative example, discussed later.

\section{Methods}

In order to resolve these issues and improve the estimation of the PIF, we propose two nonparametric methods:
one that requires individual-level exposure data, which we call the ``empirical method'', and one that only uses the mean and variance of the exposure data, which we call the ``approximate method''. Both methods are implemented in an \texttt{R} package \texttt{pifpaf}, available on Github (\texttt{github.com/colleenchan/pifpaf}).

\subsection{Empirical method}\label{point_estimate}

Let $\bm{X}_1, \bm{X}_2, \dots, \bm{X}_n$ be a random sample of $n$ individuals to whom we wish to apply the PIF or PAF.  

Denote ${\mu}^{\textrm{obs}}(\bm{\beta})$ the mean of the relative risk, conditional on $\bm{\beta}$. It can be estimated by
\begin{equation*}%\label{eq:mucftmuobs}
\widehat{\mu}_n^{\textrm{obs}}(\bm{\beta}) = \dfrac{1}{n} \sum\limits_{i=1}^{n}  RR\big( \bm{X}_i; \bm{\beta} \big).
\end{equation*}
Let $\mu^{\textrm{cft}}(\bm{\beta})$ denote the conditional mean under the counterfactual scenario. If the counterfactual exposure can be written as a function of the original exposure, $g(\bm{X_i})$, the counterfactual conditional mean is estimated by:
\begin{equation*}
  \widehat{\mu}_n^{\textrm{cft}}(\bm{\beta}) = \dfrac{1}{n} \sum_{i=1}^n   RR\big( g(\bm{X}_i); \bm{\beta} \big).
\end{equation*}
Let $\widehat{\bm{\beta}}$ be an estimate of $\bm{\beta}$ from a previous study or a meta analysis, ideally a causal estimate.
We define the empirical estimaters of PAF and PIF as:
\begin{equation}\label{pafestimate}
\widehat{\textrm{PAF}} := 1 - \dfrac{1}{\widehat{\mu}_n^{\textrm{obs}}(\widehat{\bm{\beta}})}, \qquad \textrm{and} \qquad \widehat{\textrm{PIF}} := 1 - \dfrac{\widehat\mu_n^{\textrm{cft}}(\widehat{\bm{\beta}})}{\widehat{\mu}_n^{\textrm{obs}}(\widehat{\bm{\beta}})}.
\end{equation}

The asymptotic properties of these estimators are presented in the following theorem.
\begin{theorem}
\label{thm1}
Suppose that $\widehat{\bm{\beta}}$ is a consistent and asymptotically normal estimator from an independent study.
That is, $\widehat{\bm{\beta}}\overset{p}{\longrightarrow}\bm{\beta}$ and $\sqrt{m} (\widehat{\bm{\beta}}-\bm{\beta})$ is asymptotically mean-zero multivariate normal with covariance matrix $\bm\Sigma_{\bm\beta}$, where $m$ is the sample size of the independent study estimating $\bm{\beta}$.
Assume $RR(\bm{X};\bm{\beta})$ is a differentiable function of $\bm{\beta}$, and integrable for any $\bm{\beta}$. Then $\widehat{\textrm{PAF}}$ converges in probability to $\textrm{PAF}$, and $\sqrt{n} (\widehat{\textrm{PAF}}-\textrm{PAF})$ converges toward a mean-zero normal distribution when both $m$ and $n$ approach infinity and $n/m\longrightarrow \lambda (<\infty)$. Furthermore, suppose $g(\bm{X})$ is continuous,
then $\widehat{\textrm{PIF}}$ converges in probability to $\textrm{PIF}$, and $\sqrt{n} (\widehat{\textrm{PIF}}-\textrm{PIF})$ converges toward a mean-zero normal distribution when both $m$ and $n$ approach infinity and $n/m\longrightarrow \lambda (<\infty)$. %Their covariance matrices can be estimated by the estimator described below.
\end{theorem}

The detailed proof is provided in the Appendix. We now derive the estimate of confidence intervals (CI's) for $\widehat{\textrm{PAF}}$. Given the asymptotic normality of $\hat{\mu}_n^{\textrm{obs}}(\hat{\bm{\beta}})$ as proven in \eqref{eq:together}, the variance of $\hat{\mu}_n^{\textrm{obs}}(\hat{\bm{\beta}})$ can be estimated by
\begin{equation}
\begin{aligned}
\widehat{\Var}(\hat{\mu}_n^{\textrm{obs}}(\hat{\bm{\beta}}))
&=  \frac{1}{n}\widehat{\Var}(RR( \bm{X} ;\hat{\bm\beta}))+\mathbb{E}(\nabla_{\bm\beta}RR( \bm{X};{\bm\beta}))\widehat{\Var}(\hat{\bm\beta})\mathbb{E}(\nabla_{\bm\beta}RR( \bm{X};{\bm\beta}))^T \big|_{\bm \beta = \bm{\hat{\beta}}}
\nonumber \\
&\approx \frac{1}{n}  \left(\frac{1}{n}\sum_{i=1}^n(RR(\bm{X_i};\hat{\bm\beta}))^2-\left(\hat{\mu}_n^{\textrm{obs}}(\hat{\bm{\beta}})\right)^2\right) + \\
& \qquad \left(\frac{1}{n}\sum_{i=1}^n\nabla_{\bm\beta}RR(\bm{X_i};{\bm\beta})\Big|_{\bm\beta=\hat{\bm\beta}}\right)\widehat{\Var}(\hat{\bm\beta})  \left(\frac{1}{n}\sum_{i=1}^n\nabla_{\bm\beta}RR(\bm{X_i};{\bm\beta})\Big|_{\bm\beta=\hat{\bm\beta}}\right)^T . \label{eq:muobs}
\end{aligned}
\end{equation}
where $\nabla_{\bf \beta}RR( \bm{X};{\bm\beta})$ is the gradient of $RR( \bm{X};{\bm\beta})$ with respect to $\bm \beta$.

By the delta method, the variance of $\widehat{\textrm{PAF}}$ can be estimated by
\begin{equation}
\widehat{\Var}(\widehat{\textrm{PAF}})\approx \frac{\widehat{\Var}(\hat{\mu}_n^{\textrm{obs}}(\hat{\bm{\beta}}))}{(\hat{\mu}_n^{\textrm{obs}}(\hat{\bm{\beta}}))^4}
\end{equation}
Then, the $100(1-\alpha)$\% confidence interval for $\widehat{\textrm{PAF}}$ is estimated as $\widehat{\textrm{PAF}}\pm z_{1-\frac{\alpha}{2}} \sqrt{\widehat{\Var}(\widehat{\textrm{PAF}})}$ with $z_{1-\frac{\alpha}{2}}$ the $1-\frac{\alpha}{2}$ quantile of the standard normal distribution. Similarly, the confidence intervals for $\widehat{\textrm{PIF}}$ can be constructed using the estimate of the variance of $\widehat{\textrm{PIF}}$ provided in the Appendix.

\subsection{Approximate Method}
\label{approxempirical}
Often, such as encountered by the Global Burden of Disease group, individual-level exposure data is not available \citep{forouzanfar2015global}. Rather, only the mean and variance of the exposure from a given population to which the PIF or PAF is to be applied are available. Recall that 
$\bm{X} = (X_1, X_2, \dots, X_k)^T$ where $X_1,  \dots, X_k$ are $k$ components of the exposure. Let $\bar{\bm{X}} = (\bar{X}_1, \bar{X}_2, \dots, \bar{X}_k)^T$ be the mean and $\hat{\sigma}_{i,j}$ be the covariance estimators between components $X_i$ and $X_j$. When the relative risk function $RR(\bm{X}; \bm{\beta})$ is twice differentiable in $\bm{X}$, as
%has a second order Taylor expansion for each $\bm{\beta}$ that is continuous in $\bm{\beta}$, for all $\bm{X}$ 
would be the case of linear and exponential relative risk functions,
% we show in Appendix \ref{approximate_procedure} that
by expanding the Taylor series to the second order, we can approximate $\hat{\mu}^{\textrm{obs}}(\hat{\bm{\beta}})$ by
\begin{equation}\label{varmeanonly}
\hat{\mu}^{\textrm{obs}}(\hat{\bm{\beta}}) \approx RR(\bar{\bm{X}};\hat{\bm{\beta}})+\dfrac{1}{2} \sum\limits_{i,j} \hat{\sigma}_{i,j}  \frac{\partial^2 RR\left(\bm{X};\hat{\bm{\beta}}\right)}{\partial X_i \partial X_j}\big|_{\bm{X} = \bar{\bm{X}}},
\end{equation}
leading to the estimator of the PAF
\begin{equation}\label{eq:approx_paf}
\widehat{\textrm{PAF}}
= 1-\dfrac{1} {RR(\bar{\bm{X}};\hat{\bm{\beta}})+\frac{1}{2} \sum_{i,j} \hat{\sigma}_{i,j}  \frac{\partial^2 RR\left(\bm{X};\hat{\bm{\beta}}\right)}{\partial X_i \partial X_j}\big|_{\bm{X} = \bar{\bm{X}}}}.
\end{equation}
The detailed derivation can be found in the Appendix. 
Similarly, if the counterfactual function $g(\bm{X})$ is a twice differentiable function of ${\bm{X}}$, then
\begin{equation}\label{eq:approx_pif}
\widehat{\textrm{PIF}}= 1-\dfrac{RR\big(g(\bar{\bm{X}} ),\hat{\bm{\beta}} \big) + \frac{1}{2}\sum_{i,j} \hat{\sigma}_{i,j}\frac{\partial^2 RR\left(g(\bm{X}),\hat{\bm{\beta}} \right)}{\partial X_i \partial X_j}\big|_{\bm{X} = \bar{\bm{X}}}} {RR(\bar{\bm{X}};\hat{\bm{\beta}})+\frac{1}{2} \sum_{i,j} \hat{\sigma}_{i,j}  \frac{\partial^2 RR\left(\bm{X},\hat{\bm{\beta}}\right)}{\partial X_i \partial X_j}\big|_{\bm{X} = \bar{\bm{X}}}}.
\end{equation}

For an exponential relative risk that takes the form $RR( X, \hat{ \beta}) = \exp(\hat{\beta} X)$ with $k=1$, equation \eqref{eq:approx_paf} and equation \eqref{eq:approx_pif}  simplify to
\begin{equation}\label{eq:approx_paf_simp}
\widehat{\text{PAF}}
= 1-\dfrac{1} { \exp(\hat \beta \bar X) \left(1 +\frac{1}{2}  \hat \beta ^2  \sqrt{\widehat{\Var}(X)} \right) },
\end{equation}
\begin{equation}\label{eq:approx_pif_simp}
\widehat{\text{PIF}}
= 1-\dfrac{\exp(\hat \beta g(\bar X)) \left(1 +\frac{1}{2}\sqrt{\widehat{\Var}(X)} \left(\hat \beta^2 (g'(\bar X))^2 + \hat \beta g''(\bar X) \right) \right)}
{ \exp(\hat \beta \bar X) \left(1 +\frac{1}{2}  \hat \beta ^2  \sqrt{\widehat{\Var}(X)} \right) },
\end{equation}
respectively. 
We approximate their variance using the multivariate delta method, which is derived in the Appendix. The confidence intervals are constructed similarly to the empirical method in the previous section.

% --------------------------------------------------------------------------------------------------------------------------
\section{Illustrative Example}
\label{sec:illex}

We illustrate the use of our methodology in an analysis of the impact of reductions of SSB consumption on the incidence of type 2 diabetes, and compare the results of our method with the standard method and the mixture method calculated using equation \eqref{eq:mixturePAF}\citep{kehoe2012alcohol}.
The mixture method separates out the zero values of the exposure from the positive values of the exposure and estimates the parameters of the positive values of the exposure distribution by maximum likelihood.
SSBs are drinks with added sugar including soft drinks, flavored juice drinks, sports drinks, and sweetened tea and coffee. SSB consumption has risen in many countries, most noticeably in developing countries, in recent decades, and comprises the largest source of added sugar in the U.S. diet \citep{popkin}. This is concerning since consumption of SSBs has been linked to increased risks of incidences of obesity, diabetes, and heart disease \citep{malik, hu2013, johnson, vartanian}.

The data on SSB consumption comes from ENSANUT 2016, a probabilistic national health and nutrition survey of the Mexican population gathered between May and October of 2016 \citep{gaona2018consumo}. When measuring dietary intake, researchers should be cognizant of potential measurement error, as observed intake values may not reflect actual intake values, especially in single recall surveys \citep{naska2017dietary}. In ENSANUT 2016, respondents filled out a food frequency questionnaire for the seven days prior to the interview so measurement error may be less of an issue. 
The average consumption in this data ($n = 7762$) was 1.48 servings/day, where a serving of SSB is 12 oz or 336 ml, with standard deviation 1.38 and IQR (0.56, 1.98); about 5\% of the sample had zero consumption. 
The age-adjusted relative risk of an additional serving of SSB on the incidence of type 2 diabetes in Mexico was taken from a recent meta-analysis of the existing literature, and was 1.27/serving increase of SSB with 95\% CI (1.16, 1.38) \citep{stern2019}.

\begin{figure}[!htb]
    \centering
    \includegraphics[]{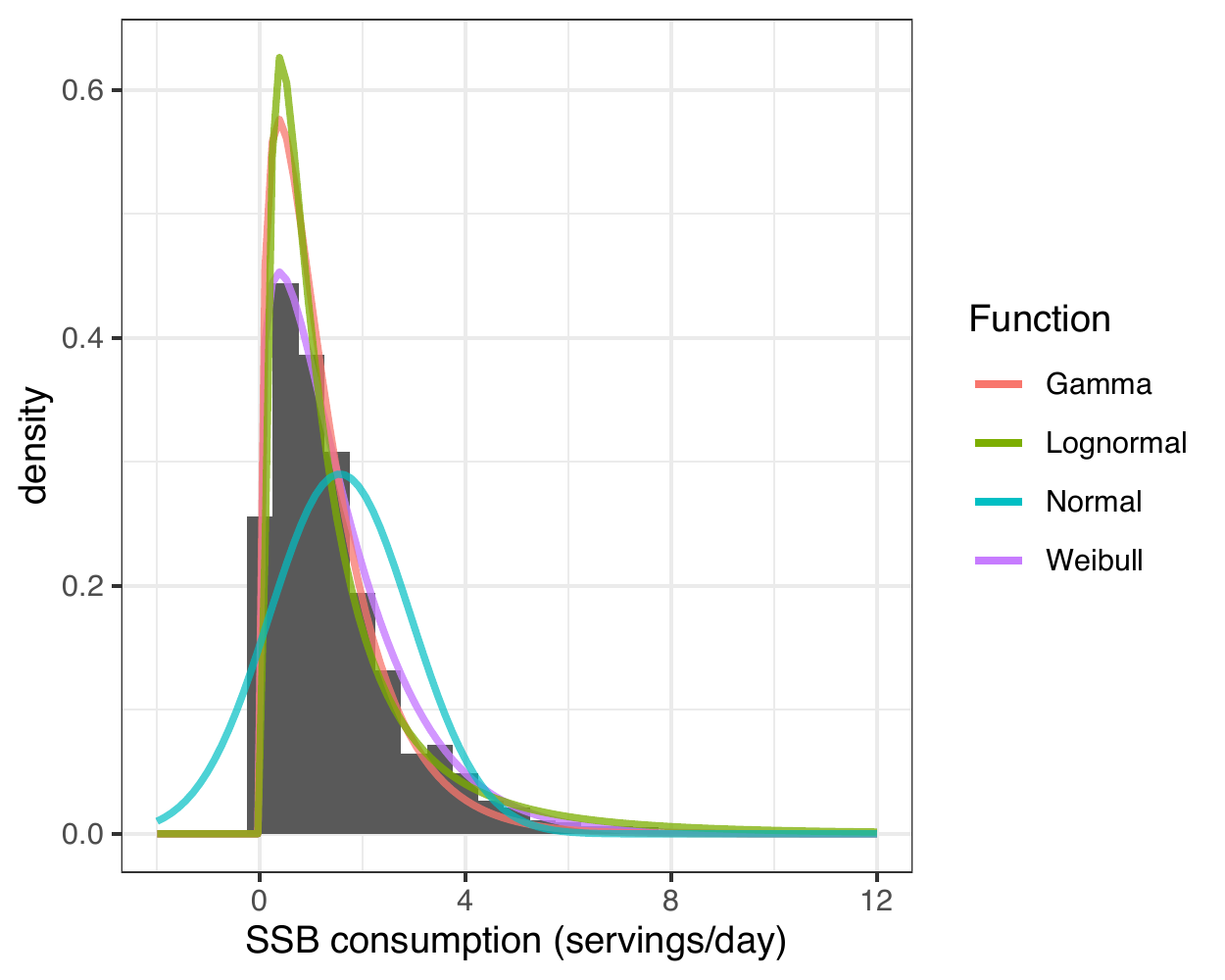} % 4in x 5in
    \caption{Distribution of SSB consumption in ENSANUT 2016.
    Gamma, Normal, Lognormal, and Weibull distributions with parameters fit using the standard method are superimposed. }
    \label{fig:ssbdensity}
\end{figure}

In the standard method, a parametric distribution, $f(x)$, must first be chosen and its parameters fit to the data using the method of moments. Then, the PAF can be estimated via equation \eqref{eq:PAF}. In the mixture method, we consider a mixture distribution where $p_0$ is the proportion of the unexposed in the data, i.e., those with zero values, and the distribution, $f(x)$, of the remaining $1-p_0$ non-zero values is fitted using maximum likelihood, as in equation \eqref{eq:mixturePAF}.  
As discussed previously, for heavy-tailed distributions, the PAF is 1 since the denominator in the second term diverges to infinity. Using a truncation bound sidesteps this issue. We estimate the PAF using the standard method and the mixture method without and with a truncation bound, where we set $M$ to be the maximum value observed in the data (11.855 servings/day). We fit several commonly considered parametric distributions, Gamma, log normal, normal, and Weibull, to the non-zero exposure values using maximum likelihood estimation. For the two-parameter Gamma and Weibull distributions, closed form solutions for the the maximum likelihood estimators are not available so the log-likelihood was maximized using the BFGS method \citep{broyden1970:BFGS}. Figure \ref{fig:ssbdensity} shows the empirical SSB consumption distribution and the fitted parametric distributions. We used Gauss-Kronrod quadrature to compute the integrations. Table \ref{tab:PAF_distr} shows the estimated PAFs using the standard method, the mixture method, and proposed empirical and approximate methods.

\begin{table}[htb!]
\begin{tabular}{lll} \toprule
  & Parameters & PAF (95\% CI) \\ \midrule
    Standard Gamma		 	 & $k = 1.15, \theta = 1.29$				&  0.345 \\
    Mixture Gamma 			 & $p_0 = 0.05, k = 1.41, \theta = 0.90$				& 0.280 \\
    Mixture Gamma ($M = 12$)     & $p_0 = 0.05, k = 1.41, \theta = 0.90$				& 0.280  \\
    Standard Lognormal 	  	 & $\log \mu = 0.08, \log \sigma = 0.31$ 	& 1     \\
    Mixture  Lognormal  		 & $p_0 = 0.05, \log \mu = 0.05, \log \sigma = 0.98$ 	& 1    \\
    Mixture Lognormal ($M = 12$) & $p_0 = 0.05, \log \mu = 0.05, \log \sigma = 0.98$  	& 0.379  \\
    Standard Normal  			 & $\mu = 1.48, \sigma = 1.38 $ 		& 0.380   \\
    Mixture  Normal 			 & $p_0 = 0.05, \mu = 1.56, \sigma =  1.37$ 		& 0.375  \\
    Mixture Normal ($M = 12$)  	 & $p_0 = 0.05, \mu = 1.56, \sigma = 1.37$ 			& 0.375  \\
    Standard  Weibull 			 & $k = 1.08, \lambda = 1.53$ 			& 0.345   \\
    Mixture Weibull   			 & $p_0 = 0.05, k = 1.20, \lambda = 1.66$ 			& 0.339   \\
    Mixture Weibull ($M$ = 12) 	 & $p_0 = 0.05, k = 1.20, \lambda = 1.66$			& 0.339  \\
    Empirical   				 & - 									& 0.345 (0.224, 0.467)   \\
    Approximate  			& - 									&  0.325 (0.219, 0.431)  \\ \bottomrule \\
\end{tabular}
    \caption{PAF estimate of type 2 diabetes due to elimination of SSB consumption in Mexico using the standard and mixture method under different distributional assumptions and our proposed empirical and approximate methods. }
    \label{tab:PAF_distr}
\end{table}

There is no method to determine when a distribution under the standard or mixture method will yield unbiased results. 
However, we note that the Weibull distribution seems to fit the SSB consumption data the best, and thus, seems to be closest to the ground truth PAF value in this application. Thus, we would hope that the PAF estimate from the empirical and approximate methods to be close to that of the Weibull estimate of approximately 0.34. This is indeed the case (0.345 for the empirical method and 0.325 for the approximate method). 

We observe that the standard method using an assumed lognormal distribution perform poorly, yielding a PAF value of 1 due to the heavy tail as aforementioned. Setting a truncation bound as in the mixture method mitigates the problem, yielding a value of 0.379, although the correct exposure distribution and truncation bound still must be properly chosen, and is typically chosen in an ad hoc manner.

% --------------------------------------------------------------------------------------------------------------------------
\section{Simulation Studies}
\label{sec:sim}

In this section, we investigated the finite sample performance of the empirical and approximate methods based on $B=10,000$ simulations, varying the sample size of the exposure $n = 100, 1000, 10000$ and the proportion of zero values $p_0 = 0, 0.05, 0.25, 0.5, 0.75$. We assumed an exponential relative risk $RR(X;\beta) = \exp(\beta X)$, where $\beta \sim Normal(\beta_0, \sigma^2(n))$ with $\beta_0 = \log(1.27)$ and $\sigma^2(n) = 70000 \cdot 0.0443^2  / (7  n) $. 
The distribution is from the estimated relative risk in our illustrative example  in Section \ref{sec:illex}, which had a sample size of 72,667 and variance $0.0443^2$. We varied the true exposure, $f(x)$, to be a truncated log normal, truncated normal, and truncated Weibull with best fit parameters also taken from our illustrative example, all truncated at $M=12$ with probability $1-p_0$ and $0$ otherwise.

For each simulation, we generated the true distribution of the exposure $f(x)$ from a mixture distribution, where we first generated $A_i, \ldots, A_n \sim \text{Bernoulli}(p_0)$. Then, we generated $X_1, \ldots, X_n \sim f(x)$ truncated at $M=12$ for $A_i > 0$ and 0 for $A_i = 0$. We also simulated $\beta \sim Normal(\beta_0, \sigma^2(n) )$, and used an exponential relative risk function $ RR(X) = \exp( \beta X)$. 
For each simulation, we estimated the PAF and the corresponding 95\% confidence interval using the empirical and approximate methods, and used the sample average and variance of the $X$ as needed. We report the coverage and average relative bias percentage in Table \ref{tab:cov} over the $B$ simulations for $f(x)$ distributed truncated Normal, truncated Lognormal, and truncated Weibull, respectively.

\begin{sidewaystable}[!htbp] \centering
\footnotesize
  \caption{
We vary the true underlying distribution $f(x)$, the proportion of 0 values $p_0$, and sample size $n$. $f(x)$ is truncated at 12 for all distributions. The parameters of each parametric distribution are taken from the illustrative example. We report the average PAF estimate, average relative bias of the PAF estimates, average PAF standard error, standard deviation of the PAF estimates, and 95\% confidence interval coverage probabilities over 10,000 simulations for each scenario.
 }
  \label{tab:cov}
\begin{tabular}{@{\extracolsep{1pt}} lllccccccccccc}
\\[-1.8ex]\toprule
& & &  \multicolumn{5}{c}{\it Empirical Method} & \multicolumn{5}{c}{\it Approximate Method} \\
\cline{5-9}\cline{10-14} \\
$f(x)$  & $p_0$ & $n$ & true PAF  & Est. & Rel. Bias  & SE & SD & Cover. & Est. & Rel. Bias & SE & SD & Cover. \\
\hline \\[-1.8ex]
Lognormal & 0.00 &   100 & 0.39 & 0.29 & -0.25 & 0.56 & 0.58 & 0.81 & 0.23 & -0.41 & 0.83 & 0.54 & 0.98 \\ 
 &  &  1000 & & 0.39 & -0.01 & 0.22 & 0.22 & 0.92 & 0.33 & -0.15 & 0.18 & 0.18 & 0.94 \\ 
 & & 10000 &  & 0.39 &  0.00 & 0.07 & 0.07 & 0.94 & 0.35 & -0.11 & 0.06 & 0.06 & 0.90 \\ 
 & 0.05 &   100 & 0.38 & 0.30 & -0.21 & 0.54 & 0.56 & 0.81 & 0.23 & -0.39 & 0.70 & 0.51 & 0.97 \\ 
 & &  1000 &  & 0.38 &  0.00 & 0.22 & 0.22 & 0.92 & 0.32 & -0.15 & 0.17 & 0.18 & 0.94 \\ 
 & & 10000 &  & 0.38 &  0.00 & 0.07 & 0.07 & 0.95 & 0.33 & -0.12 & 0.06 & 0.06 & 0.89 \\ 
 & 0.25 &  100 & 0.33 & 0.30 & -0.06 & 0.47 & 0.48 & 0.79 & 0.23 & -0.29 & 0.54 & 0.42 & 0.94 \\ 
 & &  1000 &  & 0.34 &  0.03 & 0.21 & 0.21 & 0.92 & 0.28 & -0.15 & 0.15 & 0.15 & 0.93 \\ 
 & & 10000 &  & 0.33 &  0.00 & 0.07 & 0.07 & 0.95 & 0.28 & -0.14 & 0.05 & 0.05 & 0.86 \\ 
 & 0.50 &   100 & 0.24 & 0.30 &  0.22 & 0.39 & 0.41 & 0.77 & 0.21 & -0.12 & 0.36 & 0.31 & 0.88 \\ 
 & &  1000 &  & 0.27 &  0.09 & 0.18 & 0.18 & 0.91 & 0.21 & -0.15 & 0.12 & 0.12 & 0.92 \\ 
 & & 10000 &  & 0.25 &  0.01 & 0.06 & 0.06 & 0.95 & 0.21 & -0.15 & 0.04 & 0.04 & 0.84 \\ 
 & 0.75 &   100 & 0.14 & 0.23 &  0.68 & 0.30 & 0.33 & 0.77 & 0.16 &  0.15 & 0.22 & 0.21 & 0.84 \\ 
 & &  1000 &  & 0.17 &  0.20 & 0.13 & 0.14 & 0.91 & 0.12 & -0.11 & 0.08 & 0.08 & 0.90 \\ 
 &  & 10000 &  & 0.14 &  0.02 & 0.04 & 0.04 & 0.95 & 0.12 & -0.14 & 0.03 & 0.03 & 0.86 \\ 
Normal & 0.00 &   100 & 0.36 & 0.19 & -0.48 & 0.61 & 0.67 & 0.88 & 0.18 & -0.49 & 0.70 & 0.67 & 1.00 \\ 
 &  &  1000 &  & 0.35 & -0.05 & 0.18 & 0.18 & 0.94 & 0.35 & -0.05 & 0.18 & 0.18 & 0.94 \\ 
 &  & 10000 &  & 0.36 & 0.00 & 0.06 & 0.06 & 0.95 & 0.36 & -0.01 & 0.06 & 0.06 & 0.95 \\ 
 & 0.05 &   100 & 0.35 & 0.20 & -0.43 & 0.58 & 0.61 & 0.88 & 0.20 & -0.45 & 0.66 & 0.61 & 1.00 \\ 
 &  &  1000 & & 0.34 & -0.04 & 0.18 & 0.18 & 0.94 & 0.33 & -0.05 & 0.18 & 0.18 & 0.94 \\ 
 & & 10000 & & 0.35 & 0.00 & 0.06 & 0.06 & 0.95 & 0.35 & -0.01 & 0.06 & 0.06 & 0.95 \\ 
 & 0.25 &   100 & 0.30 & 0.22 & -0.26 & 0.47 & 0.48 & 0.86 & 0.21 & -0.31 & 0.52 & 0.48 & 0.98 \\ 
 &  &  1000 &  & 0.29 & -0.03 & 0.16 & 0.16 & 0.94 & 0.29 & -0.05 & 0.16 & 0.15 & 0.94 \\ 
 & & 10000 & & 0.30 & 0.00 & 0.05 & 0.05 & 0.95 & 0.29 & -0.02 & 0.05 & 0.05 & 0.95 \\ 
 & 0.50 &   100 & 0.22 & 0.23 &  0.02 & 0.35 & 0.36 & 0.85 & 0.21 & -0.07 & 0.36 & 0.33 & 0.93 \\ 
 &  &  1000 &  & 0.22 &  0.00 & 0.13 & 0.13 & 0.94 & 0.21 & -0.04 & 0.12 & 0.12 & 0.94 \\ 
 &  & 10000 & & 0.22 &  0.00 & 0.04 & 0.04 & 0.95 & 0.22 & -0.03 & 0.04 & 0.04 & 0.94 \\ 
 & 0.75 &   100 & 0.13 & 0.18 &  0.40 & 0.25 & 0.26 & 0.85 & 0.16 &  0.26 & 0.22 & 0.21 & 0.86 \\ 
 & &  1000 &  & 0.13 &  0.05 & 0.09 & 0.09 & 0.93 & 0.13 &  0.01 & 0.08 & 0.08 & 0.93 \\ 
 & & 10000 &  & 0.13 &  0.01 & 0.03 & 0.03 & 0.95 & 0.12 & -0.01 & 0.03 & 0.03 & 0.95 \\ 
Weibull & 0.00 &   100 & 0.35 & 0.25 & -0.30 & 0.54 & 0.56 & 0.86 & 0.22 & -0.39 & 0.67 & 0.55 & 0.99 \\ 
 &  &  1000 &  & 0.34 & -0.03 & 0.19 & 0.19 & 0.94 & 0.32 & -0.08 & 0.17 & 0.17 & 0.94 \\ 
 &  & 10000 &  & 0.35 & 0.00 & 0.06 & 0.06 & 0.95 & 0.33 & -0.04 & 0.05 & 0.06 & 0.95 \\ 
& 0.05 &   100 & 0.34 & 0.25 & -0.26 & 0.52 & 0.54 & 0.86 & 0.22 & -0.36 & 0.61 & 0.52 & 0.99 \\ 
 &  &  1000 &  & 0.33 & -0.02 & 0.18 & 0.19 & 0.94 & 0.31 & -0.08 & 0.17 & 0.17 & 0.94 \\ 
 &  & 10000 &  & 0.34 & 0.00 & 0.06 & 0.06 & 0.95 & 0.32 & -0.05 & 0.05 & 0.05 & 0.94 \\ 
 & 0.25 &   100 & 0.29 & 0.26 & -0.11 & 0.44 & 0.45 & 0.84 & 0.22 & -0.25 & 0.48 & 0.42 & 0.95 \\ 
 &  &  1000 &  & 0.29 &  0.00 & 0.17 & 0.17 & 0.94 & 0.27 & -0.08 & 0.15 & 0.15 & 0.94 \\ 
 &  & 10000 &  & 0.29 & 0.00 & 0.05 & 0.05 & 0.95 & 0.27 & -0.06 & 0.05 & 0.05 & 0.94 \\ 
 & 0.50 &   100 & 0.21 & 0.25 &  0.17 & 0.36 & 0.36 & 0.83 & 0.20 & -0.04 & 0.33 & 0.30 & 0.90 \\ 
 &  &  1000 &  & 0.22 &  0.04 & 0.14 & 0.14 & 0.94 & 0.20 & -0.07 & 0.12 & 0.12 & 0.93 \\ 
 &  & 10000 &  & 0.21 &  0.01 & 0.04 & 0.05 & 0.95 & 0.20 & -0.07 & 0.04 & 0.04 & 0.93 \\ 
 & 0.75 &   100 & 0.12 & 0.19 &  0.60 & 0.26 & 0.28 & 0.83 & 0.15 &  0.28 & 0.20 & 0.20 & 0.85 \\ 
& &  1000 & & 0.13 &  0.10 & 0.09 & 0.09 & 0.93 & 0.12 & -0.02 & 0.07 & 0.07 & 0.92 \\ 
 & & 10000 &  & 0.12 &  0.01 & 0.03 & 0.03 & 0.95 & 0.11 & -0.05 & 0.02 & 0.02 & 0.94 \\ 
\bottomrule \\[-1.8ex] 
\end{tabular} 
\end{sidewaystable}

Both the approximate and empirical method performed very well, achieving minimal relative bias and excellent coverage rates.
In the simulations we considered, the empirical method maintained less than 5\% relative bias with 95\% confidence intervals achieving 93\% coverage rates when $n \ge 1,000$ in almost all scenarios and less than 1\% relative bias with approximate 95\% coverage probabilities in all cases when $n \ge 10,000$. For the empirical method estimates, the relative bias converged to 0 and the 95\% coverage probability converged to 95\% as sample size increased.  This result is expected given Theorem \ref{thm1}. 
The approximate method achieved comparable performance to the empirical method, and the relative bias decreased as the sample size increased, especially when the exposure was normal or Weibull distributed. For both methods, the average standard error of the PAF estimate and standard deviation over the $B$ simulations were very similar across all scenarios considered. 

% --------------------------------------------------------------------------------------------------------------------------

\section{Discussion}
\label{s:discuss}

The PIF is a critical epidemiological indicator, providing a primary input for disease prioritization, resource allocation and policy development. Currently, researchers rely on distributional assumptions to estimate the PIF using cross-sectional data and meta-analytic risk estimates from the literature to obtain the exposure mean and standard deviation [\cite{forouzanfar2015global}]. However, estimation methods are ill-prepared to deal with these data without making strong distributional assumptions. Moreover, to our knowledge, no methods have been developed to produce PIF estimates with individual-level data using cross-sectional surveys. Here, we characterized the implications of distributional assumptions in the estimation of the PIF and propose two nonparametric methods that overcome the observed limitations of the standard parametric methods.

The standard method for estimating the PIF is widely used to quantify the burden of disease in different countries \citep{rehm2010, forouzanfar2015global}; however, we have found at least two reasons to be cautious when implementing it. First, different distributions of exposure can lead to quite different PIF and PAF estimates, including a possible undefined result when the distribution is heavy-tailed. The problem of arbitrary selection of a parametric distribution for statistical inference have been widely discussed in the literature, e.g., \cite{wasserman2006nonparametric}. Second, truncation can also significantly bias the PIF and PAF. Only when the empirical distribution matches the selected distribution will the standard method produce an unbiased result. However, correct specification of the distribution cannot be verified. \cite{kehoe2012alcohol} compared the fit of three distributions (Weibull, Gamma and Lognormal) to fit alcohol consumption data from various countries, and concluded that the Weibull and Gamma were good fits for alcohol consumption. \cite{kehoe2012alcohol} recommended using Gamma due to its flexibility. Yet, it is unclear if the distribution of alcohol consumption is similar across countries, and there is no reason to believe that a Gamma distribution would be an adequate representation in other settings. This subjective decision-making process and the aforementioned limitations can be prevented using the nonparametric method to estimate the PIF.

Robust and nonparametric statistics avoid distribution selection problems. Nonparametric methods to estimate attributable fractions in cohort and case-control studies have been proposed \citep{wang2012comparative,hanley2001heuristic,chen2010attributable, sjolander2014doubly, taguri2012doubly}. In particular, \cite{sjolander2011estimation} and \cite{taguri2012doubly} proposed doubly robust nonparametric estimates for the PIF and PAF; however, these methods are designed for longitudinal data, where the exposure and  outcome are available from the same population. % [\cite{taguri2012doubly,sjolander2011estimation}]. 
When longitudinal data is available, using these methods will produce the best estimates. However, longitudinal data is frequently unavailable, particularly in low and middle income countries; thus, methods capable of handling exposure survey data and meta-analytical risks are needed. Our proposed methods fill this gap, allowing researchers to combine different data sources while avoiding strong distributional assumptions about the exposure.

%We have implemented these methods in the \texttt{pifpaf} package for the statistical software \textsf{R}, available at the Comprehensive R Archive Network (CRAN)\cite{pifpafpackage, rmanual}. Further information on the software's use and worked examples can be found in Appendix \ref{examples-in-r}.

Despite not requiring a specific distribution, our method is limited by standard epidemiological assumptions including transportable and unbiased relative risks, and no reverse causation \citep{rockhill1998use, basu1995cross, whittemore1982AR, bruzzi1985multiple}. In addition, assumptions about the form of the relative risk function are needed. 
For consistency, the empirical method requires that the relative risk is differentiable on $\bm{\beta}$ and is integrable with respect to the exposure distribution. 
The approximate method additionally requires that the relative risk function is twice differentiable on $\bm{X}$. These technical conditions are fulfilled by most relative risk functions and exposure distributions encountered in practice. 
Both methods require less assumptions than those used in the standard method and fare significantly better even when only the mean and variance are available, as demonstrated by our simulation studies. We have implemented these methods in the \texttt{pifpaf} package for the statistical software \textsf{R}. 

The empirical framework can be easily extended to accommodate other methods of statistical estimation. We see several avenues for future research. First, to account for outliers, robust mean estimators can be used to estimate PIF instead of the proposed $\hat{\mu}_n^{\textrm{obs}}$ \citep{Huber2011robust}. Second, nonparametric Bayesian inference is also possible, providing a compromise between an epidemiologist's conception of the data and the sample \citep{Lee2011bayes_nonparametric}. Third, our methods can be extended to adjust for measurement error in the exposure. Existing Frequentist \citep{wong2021estimation} and Bayesian inference methods \citep{chen2022bayesian} have been proposed to adjust for potential measurement error.
%Additional work is needed to advance these methods.

\backmatter

\section*{Acknowledgements}
 This work was supported by a grant from Bloomberg Philanthropies %(https://www.bloomberg.org/) 
 and the National Institute of Public Health of Mexico. TBG received support from Harvard University through the Lown Scholar's program. %(https://www.hsph.harvard.edu/lownscholars/scholars/).  
 DS was supported by a grant from the National Institutes of Health DP1ES025459.

%%%%%% include this section only if your manuscript refers to supplementary
%%%%%% materials -- see Instructions for Authors at
%%%%%% http://www.tibs.org/biometrics

%\section*{Supplementary Materials}

%Web Appendix 1 referenced in Section~\ref{ss:example} is available
%with this paper at the Biometrics website on Wiley Online Library.
%\vspace*{-8pt}

\bibliographystyle{biom} \bibliography{references}

\begin{thebibliography}{}

\bibitem[\protect\citeauthoryear{Barendregt and Veerman}{Barendregt and
  Veerman}{2010}]{barendregt2010categorical}
Barendregt, J. and Veerman, J. (2010).
\newblock Categorical versus continuous risk factors and the calculation of
  potential impact fractions.
\newblock {\em J Epidemiol Community Health} {\bf 64,} 209--212.

\bibitem[\protect\citeauthoryear{Basu and Landis}{Basu and
  Landis}{1995}]{basu1995cross}
Basu, S. and Landis, J. (1995).
\newblock Model-based estimation of population attributable risk under
  cross-sectional sampling.
\newblock {\em Am J Epidemiol} {\bf 142,} 1338--1343.

\bibitem[\protect\citeauthoryear{Broyden}{Broyden}{1970}]{broyden1970:BFGS}
Broyden, C.~G. (1970).
\newblock The convergence of a class of double-rank minimization algorithms 1.
  general considerations.
\newblock {\em IMA Journal of Applied Mathematics} {\bf 6,} 76--90.

\bibitem[\protect\citeauthoryear{Bruzzi, Green, Byar, Brinton, and
  Schairer}{Bruzzi et~al.}{1985}]{bruzzi1985multiple}
Bruzzi, P., Green, S., Byar, D., Brinton, L., and Schairer, C. (1985).
\newblock {{E}stimating the population attributable risk for multiple risk
  factors using case-control data}.
\newblock {\em Am. J. Epidemiol} {\bf 122,} 904--914.

\bibitem[\protect\citeauthoryear{Casella and Berger}{Casella and
  Berger}{2002}]{casella2002statistical}
Casella, G. and Berger, R. (2002).
\newblock {\em Statistical Inference}.
\newblock Duxbury advanced series in statistics and decision sciences. Thomson
  Learning.

\bibitem[\protect\citeauthoryear{Chen, Lin, and Zeng}{Chen
  et~al.}{2010}]{chen2010attributable}
Chen, L., Lin, D., and Zeng, D. (2010).
\newblock Attributable fraction functions for censored event times.
\newblock {\em Biometrika} {\bf 97,} 713--726.

\bibitem[\protect\citeauthoryear{Chen, Chang, Spiegelman, and Li}{Chen
  et~al.}{2022}]{chen2022bayesian}
Chen, X., Chang, J., Spiegelman, D., and Li, F. (2022).
\newblock A bayesian approach for estimating the partial potential impact
  fraction with exposure measurement error under a main study/internal
  validation design.
\newblock {\em Statistical Methods in Medical Research} {\bf 31,} 404--418.

\bibitem[\protect\citeauthoryear{Danaei, Rimm, Oza, Kulkarni, Murray, and
  Ezzati}{Danaei et~al.}{2010}]{danaei2010promise}
Danaei, G., Rimm, E., Oza, S., Kulkarni, S., Murray, C., and Ezzati, M. (2010).
\newblock The promise of prevention: the effects of four preventable risk
  factors on national life expectancy and life expectancy disparities by race
  and county in the united states.

\bibitem[\protect\citeauthoryear{Foss, Korshunov, and Zachary}{Foss
  et~al.}{2013}]{Foss2013:HeavyTail}
Foss, S., Korshunov, D., and Zachary, S. (2013).
\newblock {\em An Introduction to Heavy-Tailed and Subexponential
  Distributions}.
\newblock Springer, New York, second edition.

\bibitem[\protect\citeauthoryear{Gaona-Pineda, Mart{\'\i}nez-Tapia,
  Arango-Angarita, Valenzuela-Bravo, G{\'o}mez-Acosta, Shamah-Levy, and
  Rodr{\'\i}guez-Ram{\'\i}rez}{Gaona-Pineda et~al.}{2018}]{gaona2018consumo}
Gaona-Pineda, E.~B., Mart{\'\i}nez-Tapia, B., Arango-Angarita, A.,
  Valenzuela-Bravo, D., G{\'o}mez-Acosta, L.~M., Shamah-Levy, T., and
  Rodr{\'\i}guez-Ram{\'\i}rez, S. (2018).
\newblock Consumo de grupos de alimentos y factores sociodemogr{\'a}ficos en
  poblaci{\'o}n mexicana.
\newblock {\em salud p{\'u}blica de m{\'e}xico} {\bf 60,} 272--282.

\bibitem[\protect\citeauthoryear{GBD 2013 Risk~Factors, Forouzanfar, Alexander,
  Anderson, Bachman, Biryukov, Brauer, Burnett, Casey, Coates, and Cohen}{GBD
  2013 Risk~Factors et~al.}{2015}]{forouzanfar2015global}
GBD 2013 Risk~Factors, C., Forouzanfar, M., Alexander, L., Anderson, H.,
  Bachman, V., Biryukov, S., Brauer, M., Burnett, R., Casey, D., Coates, M.,
  and Cohen, A. (2015).
\newblock Global, regional, and national comparative risk assessment of 79
  behavioural, environmental and occupational, and metabolic risks or clusters
  of risks in 188 countries, 1990--2013: a systematic analysis for the global
  burden of disease study 2013.
\newblock {\em Lancet} {\bf 386,} 2287--2323.

\bibitem[\protect\citeauthoryear{Gmel, Shield, Frick, Kehoe, Gmel, and
  Rehm}{Gmel et~al.}{2011}]{gmel2011estimating}
Gmel, G., Shield, K., Frick, H., Kehoe, T., Gmel, G., and Rehm, J. (2011).
\newblock Estimating uncertainty of alcohol-attributable fractions for
  infectious and chronic diseases.
\newblock {\em BMC Med Res Methodol} {\bf 11,} 48.

\bibitem[\protect\citeauthoryear{Gortmaker, Long, Resch, Ward, Cradock,
  Barrett, Wright, Sonneville, Giles, Carter, Moodie, Sacks, Swinburn, Hsiao,
  Vine, Barendregt, Vos, and Wang}{Gortmaker et~al.}{2015}]{gortmaker2015cost}
Gortmaker, S., Long, M., Resch, S., Ward, Z., Cradock, A., Barrett, J., Wright,
  D., Sonneville, K., Giles, C., Carter, R., Moodie, M., Sacks, G., Swinburn,
  B., Hsiao, A., Vine, S., Barendregt, J., Vos, T., and Wang, Y. (2015).
\newblock Cost effectiveness of childhood obesity interventions: evidence and
  methods for choices.
\newblock {\em Am J Prev Med} {\bf 49,} 102--111.

\bibitem[\protect\citeauthoryear{Hanley}{Hanley}{2001}]{hanley2001heuristic}
Hanley, J. (2001).
\newblock A heuristic approach to the formulas for population attributable
  fraction.
\newblock {\em J Epidemiol Community Health} {\bf 55,} 508--514.

\bibitem[\protect\citeauthoryear{Hu}{Hu}{2013}]{hu2013}
Hu, F.~B. (2013).
\newblock Resolved: there is sufficient scientific evidence that decreasing
  sugar-sweetened beverage consumption will reduce the prevalence of obesity
  and obesity-related diseases.
\newblock {\em Obesity reviews} {\bf 14,} 606--619.

\bibitem[\protect\citeauthoryear{Huber}{Huber}{2011}]{Huber2011robust}
Huber, P. (2011).
\newblock Robust statistics.
\newblock In {\em International Encyclopedia of Statistical Science}, pages
  1248--1251. Springer Berlin Heidelberg, Berlin Heidelberg.

\bibitem[\protect\citeauthoryear{Johnson, Appel, Brands, Howard, Lefevre,
  Lustig, Sacks, Steffen, and Wylie-Rosett}{Johnson et~al.}{2009}]{johnson}
Johnson, R.~K., Appel, L.~J., Brands, M., Howard, B.~V., Lefevre, M., Lustig,
  R.~H., Sacks, F., Steffen, L.~M., and Wylie-Rosett, J. (2009).
\newblock Dietary sugars intake and cardiovascular health: a scientific
  statement from the american heart association.
\newblock {\em Circulation} {\bf 120,} 1011--1020.

\bibitem[\protect\citeauthoryear{Kehoe, Gmel, Shield, Gmel, and Rehm}{Kehoe
  et~al.}{2012}]{kehoe2012alcohol}
Kehoe, T., Gmel, G., Shield, K., Gmel, G., and Rehm, J. (2012).
\newblock Determining the best population-level alcohol consumption model and
  its impact on estimates of alcohol-attributable harms.
\newblock {\em Popul Health Metr} {\bf 10,}.

\bibitem[\protect\citeauthoryear{Lawes, Vander~Hoorn, Elliott, MacMahon, and
  Rodgers}{Lawes et~al.}{2006}]{lawes2006blood}
Lawes, C., Vander~Hoorn, Sand~Law, M., Elliott, P., MacMahon, S., and Rodgers,
  A. (2006).
\newblock Blood pressure and the global burden of disease 2000. part ii:
  estimates of attributable burden.
\newblock {\em J Hypertens} {\bf 24,} 423--430.

\bibitem[\protect\citeauthoryear{Lee}{Lee}{2011}]{Lee2011bayes_nonparametric}
Lee, J. (2011).
\newblock Bayesian nonparametric statistics.
\newblock In {\em International Encyclopedia of Statistical Science}, pages
  99--101. Springer Berlin Heidelberg, Berlin, Heidelberg.

\bibitem[\protect\citeauthoryear{Levin}{Levin}{1953}]{levin1953paf}
Levin, M. (1953).
\newblock The occurrence of lung cancer in man.
\newblock {\em Acta Unio Int Contra Cancrum} {\bf 9,} 531--541.

\bibitem[\protect\citeauthoryear{Malik, Popkin, Bray, Despr{\'e}s, Willett, and
  Hu}{Malik et~al.}{2010}]{malik}
Malik, V.~S., Popkin, B.~M., Bray, G.~A., Despr{\'e}s, J.-P., Willett, W.~C.,
  and Hu, F.~B. (2010).
\newblock Sugar-sweetened beverages and risk of metabolic syndrome and type 2
  diabetes: a meta-analysis.
\newblock {\em Diabetes care} {\bf 33,} 2477--2483.

\bibitem[\protect\citeauthoryear{Maredza, Bertram, G{\'o}mez-Oliv{\'e}, and
  Tollman}{Maredza et~al.}{2016}]{maredza2016burden}
Maredza, M., Bertram, M., G{\'o}mez-Oliv{\'e}, X., and Tollman, S. (2016).
\newblock Burden of stroke attributable to selected lifestyle risk factors in
  rural south africa.
\newblock {\em BMC Public Health} {\bf 16,} 143.

\bibitem[\protect\citeauthoryear{Murray, Ezzati, Lopez, Rodgers, and
  Vander~Hoorn}{Murray et~al.}{2003}]{murray2003comparative}
Murray, C., Ezzati, M., Lopez, A., Rodgers, A., and Vander~Hoorn, S. (2003).
\newblock Comparative quantification of health risks: conceptual framework and
  methodological issues.
\newblock {\em Popul Health Metr} {\bf 1,} 1.

\bibitem[\protect\citeauthoryear{Naska, Lagiou, and Lagiou}{Naska
  et~al.}{2017}]{naska2017dietary}
Naska, A., Lagiou, A., and Lagiou, P. (2017).
\newblock Dietary assessment methods in epidemiological research: current state
  of the art and future prospects.
\newblock {\em F1000Research} {\bf 6,}.

\bibitem[\protect\citeauthoryear{Popkin, Adair, and Ng}{Popkin
  et~al.}{2012}]{popkin}
Popkin, B.~M., Adair, L.~S., and Ng, S.~W. (2012).
\newblock Global nutrition transition and the pandemic of obesity in developing
  countries.
\newblock {\em Nutrition reviews} {\bf 70,} 3--21.

\bibitem[\protect\citeauthoryear{Rehm, Kehoe, Gmel, Stinson, Grant, and
  Gmel}{Rehm et~al.}{2010}]{rehm2010}
Rehm, J., Kehoe, T., Gmel, G., Stinson, F., Grant, B., and Gmel, G. (2010).
\newblock Statistical modeling of volume of alcohol exposure for
  epidemiological studies of population health: the us example.
\newblock {\em Population Health Metrics} {\bf 8,} 1--12.

\bibitem[\protect\citeauthoryear{Rockhill, Newman, and Weinberg}{Rockhill
  et~al.}{1998}]{rockhill1998use}
Rockhill, B., Newman, B., and Weinberg, C. (1998).
\newblock Use and misuse of population attributable fractions.
\newblock {\em Am J Public Health} {\bf 88,} 15--19.

\bibitem[\protect\citeauthoryear{Sj{\"o}lander}{Sj{\"o}lander}{2011}]{sjolander2011estimation}
Sj{\"o}lander, A. (2011).
\newblock Estimation of attributable fractions using inverse probability
  weighting.
\newblock {\em Stat Methods Med Res} {\bf 20,} 415--428.

\bibitem[\protect\citeauthoryear{Sjölander and Vansteelandt}{Sjölander and
  Vansteelandt}{2017}]{sjolander2014doubly}
Sjölander, A. and Vansteelandt, S. (2017).
\newblock Doubly robust estimation of attributable fractions in survival
  analysis.
\newblock {\em Stat Methods Med Res} {\bf 26,} 948--969.

\bibitem[\protect\citeauthoryear{Spiegelman, Hertzmark, and Wand}{Spiegelman
  et~al.}{2007}]{Spiegelman2007:PAR}
Spiegelman, D., Hertzmark, E., and Wand, H.~C. (2007).
\newblock Point and interval estimates of partial population attributable risks
  in cohort studies: examples and software.
\newblock {\em Cancer Causes Control} {\bf 18,} 571--579.

\bibitem[\protect\citeauthoryear{Stern, Mazariegos, Ortiz-Panozo, Campos,
  Malik, Lajous, and L{\'o}pez-Ridaura}{Stern et~al.}{2019}]{stern2019}
Stern, D., Mazariegos, M., Ortiz-Panozo, E., Campos, H., Malik, V.~S., Lajous,
  M., and L{\'o}pez-Ridaura, R. (2019).
\newblock Sugar-sweetened soda consumption increases diabetes risk among
  mexican women.
\newblock {\em The Journal of nutrition} {\bf 149,} 795--803.

\bibitem[\protect\citeauthoryear{Taguri, Matsuyama, Ohashi, Harada, and
  Ueshima}{Taguri et~al.}{2012}]{taguri2012doubly}
Taguri, M., Matsuyama, Y., Ohashi, Y., Harada, A., and Ueshima, H. (2012).
\newblock Doubly robust estimation of the generalized impact fraction.
\newblock {\em Biostatistics} {\bf 13,} 455--467.

\bibitem[\protect\citeauthoryear{Vander~Hoorn, Ezzati, Rodgers, Lopez, and
  Murray}{Vander~Hoorn et~al.}{2004}]{vander2004estimating}
Vander~Hoorn, S., Ezzati, M., Rodgers, A., Lopez, A., and Murray, C. (2004).
\newblock Estimating attributable burden of disease from exposure and hazard
  data.
\newblock In {\em Comparative quantification of health risks: global and
  regional burden of disease attributable to selected major risk factors},
  volume~2, pages 2129--2140. World Health Organization, Geneva.

\bibitem[\protect\citeauthoryear{Vartanian, Schwartz, and Brownell}{Vartanian
  et~al.}{2007}]{vartanian}
Vartanian, L.~R., Schwartz, M.~B., and Brownell, K.~D. (2007).
\newblock Effects of soft drink consumption on nutrition and health: a
  systematic review and meta-analysis.
\newblock {\em American journal of public health} {\bf 97,} 667--675.

\bibitem[\protect\citeauthoryear{Veerman, Sacks, Antonopoulos, and
  Martin}{Veerman et~al.}{2016}]{Veerman2016australians}
Veerman, J., Sacks, G., Antonopoulos, N., and Martin, J. (2016).
\newblock The impact of a tax on sugar-sweetened beverages on health and health
  care costs: A modelling study.

\bibitem[\protect\citeauthoryear{Walter}{Walter}{1976}]{walter1976paf}
Walter, S. (1976).
\newblock The estimation and interpretation of attributable risk in health
  research.
\newblock {\em Biometrics} {\bf 32,} 829--849.

\bibitem[\protect\citeauthoryear{Wang and Small}{Wang and
  Small}{2012}]{wang2012comparative}
Wang, W. and Small, D. (2012).
\newblock A comparative study of parametric and nonparametric estimates of the
  attributable fraction for a semi-continuous exposure.
\newblock {\em Int J Biostat} {\bf 8,} 32.

\bibitem[\protect\citeauthoryear{Wasserman}{Wasserman}{2006}]{wasserman2006nonparametric}
Wasserman, L. (2006).
\newblock {\em All of nonparametric statistics}.
\newblock Springer Science \& Business Media.

\bibitem[\protect\citeauthoryear{Whittemore}{Whittemore}{1982}]{whittemore1982AR}
Whittemore, A. (1982).
\newblock Statistical methods for estimating attributable risk from
  retrospective data.
\newblock {\em Stat Med} {\bf 1,} 229--243.

\bibitem[\protect\citeauthoryear{Wong, Lee, Spiegelman, and Wang}{Wong
  et~al.}{2021}]{wong2021estimation}
Wong, B.~H., Lee, J., Spiegelman, D., and Wang, M. (2021).
\newblock Estimation and inference for the population attributable risk in the
  presence of misclassification.
\newblock {\em Biostatistics} {\bf 22,} 805--818.

\bibitem[\protect\citeauthoryear{Young, Stensrud, Tchetgen~Tchetgen, and
  Hern{\'a}n}{Young et~al.}{2020}]{young2020causal}
Young, J.~G., Stensrud, M.~J., Tchetgen~Tchetgen, E.~J., and Hern{\'a}n, M.~A.
  (2020).
\newblock A causal framework for classical statistical estimands in
  failure-time settings with competing events.
\newblock {\em Statistics in Medicine} {\bf 39,} 1199--1236.

\end{thebibliography}

\appendix

\section{Proofs and Derivations}
\subsection{Proof of Theorem 1}

\begin{proof}
We first prove consistency of $\widehat{\textrm{PAF}}$ and $\widehat{\textrm{PIF}}$. We have,
\begin{equation*}
\hat{\mu}_n^{\textrm{obs}}(\hat{\bm{\beta}})-\mathbb{E}(RR( \bm{X} ;{\bm\beta})) = [\hat{\mu}_n^{\textrm{obs}}(\hat{\bm{\beta}})-\hat{\mu}_n^{\textrm{obs}}({\bm{\beta}})] +  [\hat{\mu}_n^{\textrm{obs}}({\bm{\beta}})-\mathbb{E}(RR(\bm{X};{\bm\beta}))]
\end{equation*}
The first term converges to zero in probability due to the consistency of $\hat{\bm{\beta}}$ and the continuous mapping theorem, and the second term converges to zero in probability due to the law of large numbers. So $\hat{\mu}_n^{\textrm{obs}}(\hat{\bm{\beta}})\overset{p}{\longrightarrow}\mathbb{E}(RR(\bm{X};{\bm\beta}))$. Similarly, $\hat\mu_n^{\textrm{cft}}(\hat{\bm{\beta}})\overset{p}{\longrightarrow}\mathbb{E}(RR(g(\bm{X});{\bm\beta}))$. The consistency of $\widehat{\textrm{PAF}}$ and $\widehat{\textrm{PIF}}$ follows directly by the continuous mapping theorem.

We then prove the asymptotic normality. By the Central Limit Theorem, conditional on $\hat{\bm\beta}$
\begin{equation*}
\sqrt{n}\left[
\left(
\begin{array}{c}
\hat{\mu}_n^{\textrm{obs}}(\hat{\bm{\beta}}) \\
\hat{\mu}_n^{\textrm{cft}}(\hat{\bm{\beta}})
\end{array}
\right)
-
\left(
\begin{array}{c}
\mathbb{E}(RR(\bm{X};\hat{\bm\beta})) \\
\mathbb{E}(RR(g(\bm{X});\hat{\bm\beta}))
\end{array}
\right)
\right]
\overset{\mathcal{D}}{\longrightarrow}
N(\bm{0},\bm\Sigma_1(\hat{\bm\beta})),
\end{equation*}
where the covariance matrix $\bm{\Sigma}_1(\cdot)$ is equal to
\begin{equation*}
\bm{\Sigma}_1(\bm\beta) = \left(
\begin{array}{cc}
\Var(RR(\bm{X};{\bm\beta})) & \Cov(RR(\bm{X};{\bm\beta}),RR(g(\bm{X});{\bm\beta})) \\
\Cov(RR(\bm{X};{\bm\beta}),RR(g(\bm{X});{\bm\beta})) & \Var(RR(g(\bm{X});{\bm\beta}))
\end{array}
\right)
\end{equation*}
By consistency of $\hat{\bm\beta}$, we have $\bm{\Sigma}_1(\hat{\bm\beta})\overset{p}{\longrightarrow}\bm{\Sigma}_1({\bm\beta})$. Then, we have
\begin{equation} \label{eq:part1clt}
\sqrt{n}\left[
\left(
\begin{array}{c}
\hat{\mu}_n^{\textrm{obs}}(\hat{\bm{\beta}}) \\
\hat{\mu}_n^{\textrm{cft}}(\hat{\bm{\beta}})
\end{array}
\right)
-
\left(
\begin{array}{c}
\mathbb{E}(RR(\bm{X};\hat{\bm\beta})) \\
\mathbb{E}(RR(g(\bm{X});\hat{\bm\beta}))
\end{array}
\right)
\right]
\overset{\mathcal{D}}{\longrightarrow}
N(\bm{0},\bm\Sigma_1(\bm \beta))
\end{equation}

By the Delta method,
\begin{equation} \label{eq:part2delta}
\sqrt{m}\left[
\left(
\begin{array}{c}
\mathbb{E}(RR(\bm{X};\hat{\bm\beta})) \\
\mathbb{E}(RR(g(\bm{X});\hat{\bm\beta}))
\end{array}
\right)
-
\left(
\begin{array}{c}
\mathbb{E}(RR(\bm{X};{\bm\beta})) \\
\mathbb{E}(RR(g(\bm{X});{\bm\beta}))
\end{array}
\right)
\right]
\overset{\mathcal{D}}{\longrightarrow}
N(\bm{0},\bm\Sigma_2),
\end{equation}
where the covariance matrix $\bm{\Sigma}_2$ is equal to
\begin{equation*}
\bm\Sigma_2 =
\left(
\begin{array}{c}
{\mathbb{E}(\nabla_{\bm\beta}RR(\bm{X};{\bm\beta}))} \\
{\mathbb{E}(\nabla_{\bm\beta}RR(g(\bm{X});{\bm\beta}))}
\end{array}
\right)
\bm\Sigma_{\bm\beta}
\left(
\begin{array}{c}
{\mathbb{E}(\nabla_{\bm\beta}RR(\bm{X};{\bm\beta}))} \\
{\mathbb{E}(\nabla_{\bm\beta}RR(g(\bm{X});{\bm\beta}))}
\end{array}
\right)^T.
\end{equation*}
Notice that
\begin{eqnarray*}
&&
\left(
\begin{array}{c}
\hat{\mu}_n^{\textrm{obs}}(\hat{\bm{\beta}}) \\
\hat{\mu}_n^{\textrm{cft}}(\hat{\bm{\beta}})
\end{array}
\right)
-
\left(
\begin{array}{c}
\mathbb{E}(RR(\bm{X};{\bm\beta})) \\
\mathbb{E}(RR(g(\bm{X});{\bm\beta}))
\end{array}
\right) \\
&=&
\underbrace{
%\left\{
\left(
\begin{array}{c}
\hat{\mu}_n^{\textrm{obs}}(\hat{\bm{\beta}}) \\
\hat{\mu}_n^{\textrm{cft}}(\hat{\bm{\beta}})
\end{array}
\right)
-
\left(
\begin{array}{c}
\mathbb{E}(RR(\bm{X};\hat{\bm\beta})) \\
\mathbb{E}(RR(g(\bm{X});\hat{\bm\beta}))
\end{array}
\right)
%\right\}
}_{\textrm{first term}}
+
\underbrace{
%\left\{
\left(
\begin{array}{c}
\mathbb{E}(RR(\bm{X};\hat{\bm\beta})) \\
\mathbb{E}(RR(g(\bm{X});\hat{\bm\beta}))
\end{array}
\right)
-
\left(
\begin{array}{c}
\mathbb{E}(RR(\bm{X};{\bm\beta})) \\
\mathbb{E}(RR(g(\bm{X});{\bm\beta}))
\end{array}
\right)
%\right\}
}_{\textrm{second term}}.
\end{eqnarray*}
The two terms are asymptotically unrelated since, by a double expectation argument,
\small
\begin{eqnarray*}
&&
\mathbb{E}\left[
\left\{
\left(
\begin{array}{c}
\hat{\mu}_n^{\textrm{obs}}(\hat{\bm{\beta}}) \\
\hat{\mu}_n^{\textrm{cft}}(\hat{\bm{\beta}})
\end{array}
\right)
-
\left(
\begin{array}{c}
\mathbb{E}(RR(\bm{X};\hat{\bm\beta})) \\
\mathbb{E}(RR(g(\bm{X});\hat{\bm\beta}))
\end{array}
\right)
\right\}^T
\left\{
\left(
\begin{array}{c}
\mathbb{E}(RR(\bm{X};\hat{\bm\beta})) \\
\mathbb{E}(RR(g(\bm{X});\hat{\bm\beta}))
\end{array}
\right)
-
\left(
\begin{array}{c}
\mathbb{E}(RR(\bm{X};{\bm\beta})) \\
\mathbb{E}(RR(g(\bm{X});{\bm\beta}))
\end{array}
\right)
\right\}
\right] \\
&=&
\mathbb{E}\left[
\mathbb{E}
\left\{
\left(
\begin{array}{c}
\hat{\mu}_n^{\textrm{obs}}(\hat{\bm{\beta}}) \\
\hat{\mu}_n^{\textrm{cft}}(\hat{\bm{\beta}})
\end{array}
\right)
-
\left(
\begin{array}{c}
\mathbb{E}(RR(\bm{X};\hat{\bm\beta})) \\
\mathbb{E}(RR(g(\bm{X});\hat{\bm\beta}))
\end{array}
\right)
\Big|
\hat{\bm\beta}
\right\}^T
\left\{
\left(
\begin{array}{c}
\mathbb{E}(RR(\bm{X};\hat{\bm\beta})) \\
\mathbb{E}(RR(g(\bm{X});\hat{\bm\beta}))
\end{array}
\right)
-
\left(
\begin{array}{c}
\mathbb{E}(RR(\bm{X};{\bm\beta})) \\
\mathbb{E}(RR(g(\bm{X});{\bm\beta}))
\end{array}
\right)
\right\}
\right]\\
&\longrightarrow&0.
\end{eqnarray*}

\normalsize

Then, we have
\begin{equation} \label{eq:together}
\sqrt{n}\left[
\left(
\begin{array}{c}
\hat{\mu}_n^{\textrm{obs}}(\hat{\bm{\beta}}) \\
\hat{\mu}_n^{\textrm{cft}}(\hat{\bm{\beta}})
\end{array}
\right)
-
\left(
\begin{array}{c}
\mathbb{E}(RR(\bm{X};{\bm\beta})) \\
\mathbb{E}(RR(g(\bm{X});{\bm\beta}))
\end{array}
\right)
\right]
\overset{\mathcal{D}}{\longrightarrow}
N(\bm{0},\bm\Sigma),
\end{equation}
where $\bm\Sigma=\bm\Sigma_1(\bm\beta)+ \bm\Sigma_2$.

Asymptotic normality is obtained by the delta method. That is,
\begin{equation*}
\sqrt{n}(\widehat{\textrm{PAF}}-\textrm{PAF}) \overset{D}{\longrightarrow}N(0,\sigma_{PAF}^2),
\end{equation*}
where $\sigma_{PAF}^2=\Sigma_{11}/\mathbb{E}(RR(\bm{X};{\bm\beta}))^4$ and $\Sigma_{11}$ is the first diagonal entry of $\bm\Sigma$,
\begin{equation*}
\Sigma_{11} = \Var(RR(\bm{X};{\bm\beta}))+ \mathbb{E}(\nabla_{\bm\beta}RR(\bm{X};{\bm\beta}))\bm\Sigma_{\bm\beta}\mathbb{E}(\nabla_{\bm\beta}RR(\bm{X};{\bm\beta}))^T,
\end{equation*}

and
\begin{equation*}
\sqrt{n}(\widehat{\textrm{PIF}}-\textrm{PIF}) \overset{D}{\longrightarrow}N(0,\sigma_{PIF}^2),
\end{equation*}
where the asymptotic variance
\begin{equation*}
\sigma_{PIF}^2 =
\left(
\begin{array}{cc}
\frac{\mathbb{E}(RR(g(\bm{X});{\bm\beta}))}{\mathbb{E}(RR(\bm{X};{\bm\beta}))^2} & -\frac{1}{\mathbb{E}(RR(\bm{X};{\bm\beta}))}
\end{array}
\right)
\bm\Sigma
\left(
\begin{array}{cc}
\frac{\mathbb{E}(RR(g(\bm{X});{\bm\beta}))}{\mathbb{E}(RR(\bm{X};{\bm\beta}))^2} & -\frac{1}{\mathbb{E}(RR(\bm{X};{\bm\beta}))}
\end{array}
\right)^T.
\end{equation*}
\end{proof}

\subsection{Approximate Method Estimator}
The first and second moments of $\bm{X}$ are
\begin{equation*}
\bm\mu_{\bm{X}} = \mathbb{E}(\bm{X}) \quad \textrm{ and } \quad {\bm\Sigma}_{\bm{X}} = \Var(\bm{X}),
\end{equation*}
and their estimates are
\begin{equation*}
\widehat{\bm\mu}_{\bm{X}} = \frac{1}{n}\sum_{i=1}^n\bm{X}_i = \bar{\bm{X}} \quad \textrm{ and } \quad \widehat{\bm\Sigma}_{\bm{X}} = \frac{1}{n}\sum_{i=1}^n(\bm{X}_i-\widehat{\bm\mu}_{\bm{X}})(\bm{X}_i-\widehat{\bm\mu}_{\bm{X}})^T.
\end{equation*}

Consider a general function $h(\bm{X})$, which is twice differentiable. Let
\begin{equation*}
\bm{D}h(\bm{X}) = \frac{\partial h(\bm{X})}{\partial \bm{X}} \quad \textrm{ and } \quad \bm{H}h(\bm{X}) = \frac{\partial^2 h(\bm{X})}{\partial \bm{X} \partial \bm{X}^T}.
\end{equation*}
The second-order Taylor polynomial for $h(\bm{X})$ is
\begin{eqnarray*}
h(\bm{X}) & \approx & h(\widehat{\bm\mu}_{\bm{X}}) + \bm{D}h(\widehat{\bm\mu}_{\bm{X}})(\bm{X}-\widehat{\bm\mu}_{\bm{X}})+\frac{1}{2}(\bm{X}-\widehat{\bm\mu}_{\bm{X}})^T\bm{H}h(\widehat{\bm\mu}_{\bm{X}})(\bm{X}-\widehat{\bm\mu}_{\bm{X}}) \\
& = & h(\widehat{\bm\mu}_{\bm{X}}) + \bm{D}h(\widehat{\bm\mu}_{\bm{X}})(\bm{X}-\widehat{\bm\mu}_{\bm{X}})+\frac{1}{2}tr\left[(\bm{X}-\widehat{\bm\mu}_{\bm{X}})(\bm{X}-\widehat{\bm\mu}_{\bm{X}})^T\bm{H}h(\widehat{\bm\mu}_{\bm{X}})\right].
\end{eqnarray*}
So applying the approximation to all subjects $\bm{X}_1$,  $\bm{X}_2$, ..., $\bm{X}_n$, we have,
\begin{equation*}
\frac{1}{n}\sum_{i=1}^nh(\bm{X}_i) \approx h(\widehat{\bm\mu}_{\bm{X}}) +  \frac{1}{2}tr\left[\widehat{\bm\Sigma}_{\bm{X}}\bm{H}h(\widehat{\bm\mu}_{\bm{X}})\right].
\end{equation*}

Using this, we can approximate $\hat{\mu}_n^{\textrm{obs}}(\hat{\bm{\beta}})$ and $\hat{\mu}_n^{\textrm{cft}}(\hat{\bm{\beta}})$ as

\begin{equation}
\hat{\mu}^{\textrm{obs}}(\hat{\bm{\beta}}) \approx RR(\bar{\bm{X}};\hat{\bm{\beta}})+\dfrac{1}{2} \sum\limits_{i,j} \hat{\sigma}_{i,j}  \frac{\partial^2 RR\left(\bm{X};\hat{\bm{\beta}}\right)}{\partial X_i \partial X_j}\big|_{\bm{X} = \bar{\bm{X}}},
\end{equation}
\begin{equation}
\hat{\mu}^{\textrm{cft}}(\hat{\bm{\beta}}) \approx RR\big(g(\bar{\bm{X}} ),\hat{\bm{\beta}} \big) + \frac{1}{2}\sum_{i,j} \hat{\sigma}_{i,j}\frac{\partial^2 RR\left(g(\bm{X}),\hat{\bm{\beta}} \right)}{\partial X_i \partial X_j}\big|_{\bm{X} = \bar{\bm{X}}}.
\end{equation}
The approximate method PIF and PAF estimators follow directly by substituting these quantities into equation \eqref{pafestimate}.

\subsection{Variance of the Approximate Method Estimator}
Note that the PAF estimator in equation \eqref{eq:approx_paf_simp} can be expressed as a function of three parameters: \[ \widehat{ \text{PAF}} = h( \bm Z) = h(\bm{\bar X}, \widehat{\text{Var}}(\bm X),  \hat{\bm{\beta}} ) \]

If $Z$ is a consistent estimator for $\zeta$, then we can use multivariate delta method to obtain asymptotic normality
\[
\sqrt{n} (h( \bm Z) - h( \bm \zeta)) \rightarrow \mathcal{N}(0, \nabla h( \bm Z)^T \cdot  \bm \Sigma \cdot \nabla h(\bm Z))
\]
where
\[ \nabla h(\bm{Z})^T =
\begin{pmatrix}
 \frac{ \hat{\bm{\beta}}}{\exp(\hat{\bm{\beta}} \bm{\bar X}) \left(1 + \frac12  \hat{\bm{\beta}}^2 \sqrt{\Var(\bm{X}) } \right) } &
 \frac{ \hat \beta^2 \exp (\hat{\bm{\beta}} \bar{\bm X}) \left( 1 + \frac12 \hat{\bm{\beta}}^2 \sqrt{\Var(\bm{X}) } \right)^2}{ 4 \sqrt{\Var(\bm{X})} } &
 \frac{ \bar{\bm{X}} + \hat{\bm{\beta}} (1 + \frac12 \hat{\bm{\beta}}) \sqrt{\Var(\bm{X})} }{ \exp( \hat{\bm{\beta}} \bar{\bm X}) \left( 1 + \frac12 \hat{\bm{\beta}}^2 \sqrt{\Var (\bm{X})} \right)^2 }
\end{pmatrix},
 \]
 and
\[
\begin{aligned}
 \bm \Sigma &=
\begin{pmatrix}
\Var(\bar{\bm{X}})			 			& \Cov(\bar{\bm{X}}, \widehat{\Var}(\bm{X}))		& \Cov(\bar{\bm{X}}, \hat{\bm{\beta}}) \\
\Cov(\bar{\bm{X}}, \widehat{\Var}(\bm{X}))  	& \Var( \widehat{\Var}(\bm{X})) 				& \Cov (\widehat{\Var}(\bm{X}), \hat{\bm{\beta}}) \\
\Cov(\bar{\bm{X}}, \hat{\bm{\beta}}	)			& \Cov (\widehat{\Var}(\bm{X}), \hat{\bm{\beta}})	& \Var( \hat{\bm{\beta}})
\end{pmatrix} \\
%&=
%\begin{pmatrix}
%\Var(\bar{X})		& 0								& 0 \\
%0  					& \Var( \widehat{\Var}(X)) 		& 0 \\
%0					& 0								& \Var( \hat \beta)
%\end{pmatrix} \\
&\approx
\begin{pmatrix}
\frac{\widehat{\Var}(\bm{X})}{n}		& 0								& 0 \\
0  					&  \frac{3\widehat{\Var}(\bm{X})^2}{n} - \frac{(n-3)\widehat{\Var}(\bm{X})^{3/2}}{n(n-1)}		& 0 \\
0					& 0								& \Var( \hat{\bm{\beta}})
\end{pmatrix}.
\end{aligned}
\]
The covariance terms $\Cov(\bar{\bm{X}}, \hat{\bm{\beta}})$ and $\Cov (\widehat{\Var}(\bm{X}), \hat{\bm{\beta}})$ are 0 since they are taken from independent studies. If $\bm{X}$ is normally distributed, $\Cov(\bar{ \bm X}, \widehat{\Var}(\bm{X})) = 0$ and  $\Var( \widehat{\Var({\bm{X})}})
= \frac{ 3 \Var(\bm{X})^2 }{n} - \frac{ \Var(\bm{X})^{3/2} (n-3)}{n(n-1)}$. So we approximate $\Cov(\bar{\bm X}, \widehat{\Var}( \bm X)) \approx 0$ and $\Var( \widehat{\Var}(\bm{X}))
\approx \frac{ 3 \widehat{\Var}(\bm{X})^2 }{n} - \frac{ \widehat{\Var}(\bm{X})^{3/2} (n-3)}{n(n-1)}$.

$\widehat \Var(\text{PAF}) = \nabla h( \bm Z)^T \cdot  \bm \Sigma \cdot \nabla h(\bm Z)$ is used as the estimate of variance of PAF for the approximate method. Similarly, we can derive the variance of the PIF for the approximate method.

\label{lastpage}

\end{document}